\documentclass[aps,reprint,amsmath,amssymb,floatfix,superscriptaddress,longbibliography]{revtex4-2}

\usepackage[english]{babel}
\usepackage{soul}
\usepackage{graphicx}
\usepackage{dcolumn}
\usepackage{bm}
\usepackage[per-mode=symbol]{siunitx}

\DeclareSIUnit{\bits}{bits}
\DeclareSIUnit{\wtpercent}{wt\%}
\usepackage[T1]{fontenc}
\usepackage{mathptmx}
\usepackage{makecell}
\usepackage{threeparttable}
\usepackage{xcolor}
\usepackage{nicefrac}

\usepackage{braket}

\usepackage{tikz}
\usepackage{array}
\usepackage{tabularx}
\usepackage{ragged2e}
\usepackage{hyperref}

\definecolor{hotblue}{RGB}{46,48,146}
\hypersetup{
            colorlinks=true,
            linkcolor=hotblue,
            citecolor=hotblue,
            urlcolor=hotblue,
            }
\definecolor{blue}{RGB}{97,111,183}
\definecolor{orange}{RGB}{239,134,54}
\definecolor{green}{RGB}{76,167,96}
\definecolor{red}{RGB}{235,127,126}
\definecolor{yellow}{RGB}{235,235,100}


\usepackage{microtype}
\makeatletter
\renewcommand{\fnum@figure}{\textbf{Fig.~\thefigure}}
\renewcommand{\@caption@fignum@sep}{\enspace|\enspace}
\makeatother

\begin{document}

\title{Cooling-path-selected self-assembly of telecom InAsP quantum dots}

\author{Yiteng Zhang}
\author{Doaa Abdelbarey}
\affiliation{Institute of Solid State Physics, Leibniz University Hannover, Appelstra{\ss}e 2, 30167 Hannover, Germany}

\author{Markus Etzkorn}
\affiliation{Technische Universität Braunschweig, Laboratory for Emerging Nanometrology (LENA), Langer Kamp 6a, 38106 Braunschweig, Germany}

\author{Zenghui Jiang}
\author{Ankita Choudhary}
\affiliation{Institute of Solid State Physics, Leibniz University Hannover, Appelstra{\ss}e 2, 30167 Hannover, Germany}

\author{Andreas Schaate}
\affiliation{Institute of Inorganic Chemistry, Leibniz Universität Hannover, Callinstraße 3–9, 30167 Hannover, Germany}

\author{Tom Fandrich}
\author{Xin Cao}
\author{Chenxi Ma}
\author{Arijit Chakraborty}
\affiliation{Institute of Solid State Physics, Leibniz University Hannover, Appelstra{\ss}e 2, 30167 Hannover, Germany}

\author{Eddy Patrick Rugeramigabo}
\affiliation{Institute of Solid State Physics, Leibniz University Hannover, Appelstra{\ss}e 2, 30167 Hannover, Germany}

\author{Michael Zopf}
\email{michael.zopf@fkp.uni-hannover.de}
\affiliation{Institute of Solid State Physics, Leibniz University Hannover, Appelstra{\ss}e 2, 30167 Hannover, Germany}
\affiliation{Laboratory of Nano and Quantum Engineering, Leibniz University Hannover, Schneiderberg 39, 30167 Hannover, Germany}

\author{Fei Ding}
\email{fei.ding@fkp.uni-hannover.de}
\affiliation{Institute of Solid State Physics, Leibniz University Hannover, Appelstra{\ss}e 2, 30167 Hannover, Germany}
\affiliation{Laboratory of Nano and Quantum Engineering, Leibniz University Hannover, Schneiderberg 39, 30167 Hannover, Germany}

\begin{abstract}
Epitaxial quantum dots are leading solid-state sources of quantum light, but reproducible growth on indium phosphide (InP) remains challenging because dot-formation windows are narrow and emission wavelengths are difficult to predict. Here we report a new epitaxial route for InAsP quantum dots on InP(001), in which the cooling path selects the final morphology without intentional deposition of group-III dot material. High-temperature As exposure forms a shallow near-surface InAsP layer through As--P exchange; the cooling path and continued As supply determine whether three-dimensional islands form. Surface-sensitive spectroscopy shows near-surface As enrichment, while cross-sectional STEM/EELS identifies graded, As-rich InAsP islands. The capped emitters span 1.2--1.6~\textmu m, with single-photon emission observed up to 100~K. Across 20 wafers, the surface-effective exchange dose $D$ provides a practical calibration for targeting the median emission wavelength. These results establish cooling as an active epitaxial step for telecom InAsP quantum-dot formation and wavelength control.
\end{abstract}

\maketitle
Epitaxial quantum dots (QDs) are leading solid-state sources of on-demand quantum light. They generate highly indistinguishable single photons and entangled photon pairs with high efficiency \cite{Somaschi2016NearOptimal,Wang2019OptimalSPS,Liu2019EntangledPhotonPairs,Rota2024EntangledQD}. They have enabled cluster-state generation \cite{Cogan2023ClusterState} and scalable photonic quantum-information architectures \cite{Wang2020IntegratedPhotonicTechnologies,Maring2024SinglePhotonQC,Chen2021PhotonicsPerspective}. Their epitaxial nature allows the emission wavelength and device structure to be engineered for low-loss telecom bands and photonic integration \cite{Wang2025ScalablePhotonic,Yu2023TelecomReview}. Recent work has integrated QD sources with electrically injected pump lasers, CMOS-compatible photonic circuits and site-controlled cavity arrays \cite{Li2023IntegratedMicrolaser,Zhu2025HybridQDSiC,Gaur2026ScalableSiteControlledQD}. At telecom wavelengths, QD devices have demonstrated coherent and indistinguishable single photons \cite{Albrechtsen2026QuantumCoherent,Hauser2026TelecomIndistinguishable,Holewa2024TelecomSingleEmitter}, entangled photon pairs \cite{Muller2018TelecomQLED}, spin--photon entanglement \cite{Laccotripes2024SpinPhotonTelecom}, quantum teleportation \cite{Anderson2020QuantumTeleportation} and intercity quantum key distribution \cite{Yang2024IntercityQKD}. Moving beyond isolated single-emitter demonstrations requires reproducible epitaxy, predictable wavelength control and compatibility with wafer-scale processing.

Telecom QDs have been realized through several epitaxial routes in III--V systems. In Stranski--Krastanov (SK) growth, deposition first forms a strained wetting layer, and further deposition beyond a critical thickness produces three-dimensional islands as strain relaxes \cite{Tersoff1993,Shchukin1999Spontaneous}. This route has produced high-quality InAs/GaAs QDs in the near-infrared, but extending their emission to telecom wavelengths generally requires additional strain or composition engineering \cite{Kolatschek2021BrightPurcellOband,Nawrath2023BrightCband}. Strain-reducing capping layers and indium-flush techniques provide further control of QD shape and emission wavelength \cite{Spitzer2024TelecomOBandLDE,Sasakura2007IndiumFlush}. In droplet-based routes, group-III droplets can be crystallized under a group-V flux to form QDs or used to etch nanoholes before QD material is deposited. With nanohole infill, QD size and emission wavelength can be tuned through the amount and composition of the deposited material \cite{Gurioli2019DropletReview,Spitzer2024TelecomOBandLDE,Polczynska2026LDETelecomCband}. Recent InAs/GaAs QDs formed in droplet-etched nanoholes have reached the telecom O-band with an additional InGaAs strain-reduction layer \cite{Spitzer2024TelecomOBandLDE}. GaSb/AlGaSb QDs formed by local droplet etching (LDE) have also shown single-photon emission in the telecom S-band \cite{Michl2023StrainFreeGaSb}.

InP supports the monolithic integration of lasers, optical amplifiers and modulators \cite{Smit2019InPIntegration}. InAs/InP QDs can also emit directly at telecom wavelengths \cite{miyazawa_single-photon_2016,Holewa2024TelecomSingleEmitter}. Yet reliable QD growth on InP remains difficult. The smaller lattice mismatch of InAs/InP (\(\sim\)3.2\%) relative to InAs/GaAs (\(\sim\)7\%) reduces the strain energy available to drive three-dimensional islanding \cite{Shchukin1999Spontaneous}. InAs growth often produces elongated quantum dashes or rods rather than compact dot-like structures \cite{Gendry2004QuantumSticks,Dery2004QuantumDashNature}, and the final morphology is highly sensitive to temperature, group-V conditions and substrate orientation \cite{gutierrez_role_2002,Bierwagen2005VicinalInP}. Droplet-based InP routes have produced high-quality telecom emitters through droplet epitaxy or LDE followed by nanohole filling \cite{Gurioli2019DropletReview,Holewa2022DropletInAsInP,Polczynska2026LDETelecomCband}. The LDE route has recently achieved C-band single-photon emission but requires separate control of nanohole formation, filling and overgrowth \cite{Polczynska2026LDETelecomCband}. A remaining challenge is to establish a reproducible InP route that combines reliable QD formation with predictable emission wavelengths without separate steps for droplet formation, nanohole etching or nanohole filling.

In molecular-beam epitaxy (MBE), As exposure during oxide removal and surface stabilization can drive As--P exchange at the InP surface. On InP(100), high-temperature As exposure can replace P within the top few surface bilayers, producing an ultrathin InAs-like or InAsP surface layer and As-rich surface reconstructions \cite{averbeck_oxide_1991,Kozhukhov1997InPAs4,Li2002AsExchangeInP}. Under specific exchange conditions, three-dimensional InAs or As-rich InAsP islands have also been reported \cite{Wang1998AsPExchangeInAsQD,kolosovsky_determination_2022}. For As-treated InP without intentional InAs deposition, As--P exchange during cooling was reported to contribute to island growth, and the resulting structures showed telecom-range ensemble photoluminescence \cite{Wang1998AsPExchangeInAsQD}. Related work on intentionally deposited InAs on InP-related surfaces showed that As--P exchange and post-deposition cooling or ripening can modify dot formation, island morphology and emission wavelength \cite{Yoon1999AsPExchangeInAsInP,Taskinen1997InAsInPIslands,gutierrez_role_2002,leavitt_pathway_2015,Yacob2014Ripening,Kors2018Ripening}. More recently, AsH$_3$ annealing in metal-organic vapor-phase epitaxy (MOVPE) produced tunable two-dimensional In(As,P) layers, although further InAs deposition was still required to form QDs \cite{Berdnikov2024}. However, these studies did not establish how cooling history and group-V conditions determine whether an exchange-generated near-surface InAsP layer develops into large three-dimensional islands, or demonstrate a reproducible route to telecom single-photon emitters.

Here we show a cooling-path-selected self-assembly (CPSA) route for telecom InAsP QDs. High-temperature As exposure first forms a shallow near-surface InAsP layer without intentional deposition of group-III dot material. The subsequent cooling path determines whether this layer develops into three-dimensional islands. The resulting capped emitters show telecom single-photon emission up to 100~K. Angle-resolved X-ray photoelectron spectroscopy (ARXPS) reveals near-surface As enrichment across the cooling-path controls. High-angle annular dark-field scanning transmission electron microscopy (HAADF-STEM) and electron energy-loss spectroscopy (EELS) identify graded, As-rich InAsP islands. Across 20 wafers grown with As-on cooling, a surface-effective exchange dose $D$ provides a practical wafer-level calibration of the median emission wavelength. The corresponding emitter populations span 1.2--1.6~\textmu m.

\section{Results}

\begin{figure}[!ht]
\centering
\includegraphics[width=0.98\linewidth]{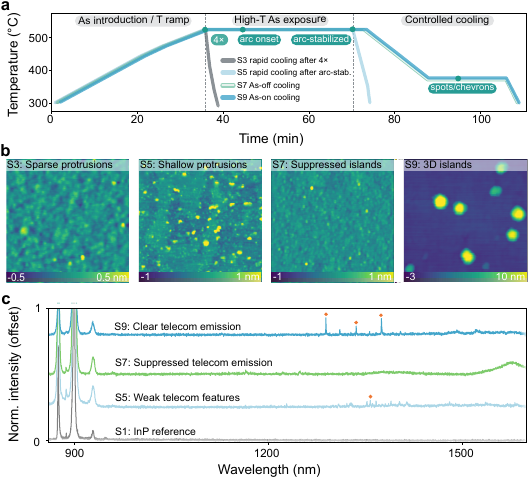}
\caption{\textbf{Cooling-path control of island formation and telecom emission on InP(001).}
\textbf{a} Logged MBE temperature--time profiles for the main control paths. Solid green dots mark four labeled stages in the reflection high-energy electron diffraction (RHEED) evolution: appearance of the streaky $4\times$ pattern, arc onset, the arc-stabilized stage and the later spotty/chevron-like pattern during cooling. Arc onset denotes the first appearance of arc-like RHEED features, whereas arc-stabilized denotes the later high-temperature stage at which the arc-like pattern changes little under continued As exposure. S3 is rapidly cooled after the $4\times$ stage; S5 is rapidly cooled with the As flux maintained after the arc-stabilized stage; S7 and S9 are cooled from the same stage with the As flux off and on, respectively. The full growth matrix and RHEED sequences are provided in Supplementary Note~1 and Supplementary Fig.~1.
\textbf{b} \textit{Ex situ} atomic force microscopy (AFM) height maps of the uncapped surfaces obtained after the four paths in \textbf{a}. Each image covers an area of $1\times1~\mu\mathrm{m}^{2}$; color scales are set independently to the height range of each surface.
\textbf{c} Low-temperature $\mu$PL spectra of the InP reference and capped samples prepared under the corresponding process conditions; solid orange dots mark representative telecom emission peaks.}
\label{Figure1}
\end{figure}

\subsection{Cooling-path control of island formation}
We first used growth-interruption and As-on/As-off cooling controls to distinguish high-temperature surface changes from the morphology that develops during cooling (Fig.~\ref{Figure1}a). During the common sequence, \textit{in situ} reflection high-energy electron diffraction (RHEED) evolves from a streaky $4\times$ pattern through arc onset to the arc-stabilized stage. We then initiated rapid cooling at selected stages, or continued slow cooling with the As shutter open or closed.

Atomic force microscopy (AFM) reveals distinct uncapped surface morphologies (Fig.~\ref{Figure1}b). Rapid cooling after the $4\times$ stage (S3) leaves a nearly flat surface with only sparse protrusions. Rapid cooling after arc onset (S4; Supplementary Fig.~2) and after the arc-stabilized stage (S5) produces sparse features and shallow protrusions, respectively, but not the large three-dimensional islands seen in S9. Starting from the same arc-stabilized stage, As-off cooling (S7) suppresses island formation, whereas slow As-on cooling followed by the low-temperature hold (S9) produces large three-dimensional islands. Together, these controls show that the high-temperature sequence does not fix the final island morphology. Within the tested paths, large islands develop only during slow As-on cooling and the subsequent low-temperature hold; closing the As shutter suppresses their formation. The full S1--S9 AFM series and morphology statistics are summarized in Supplementary Fig.~2.

\begin{figure*}[!ht]
\centering
\includegraphics[width=0.98\linewidth]{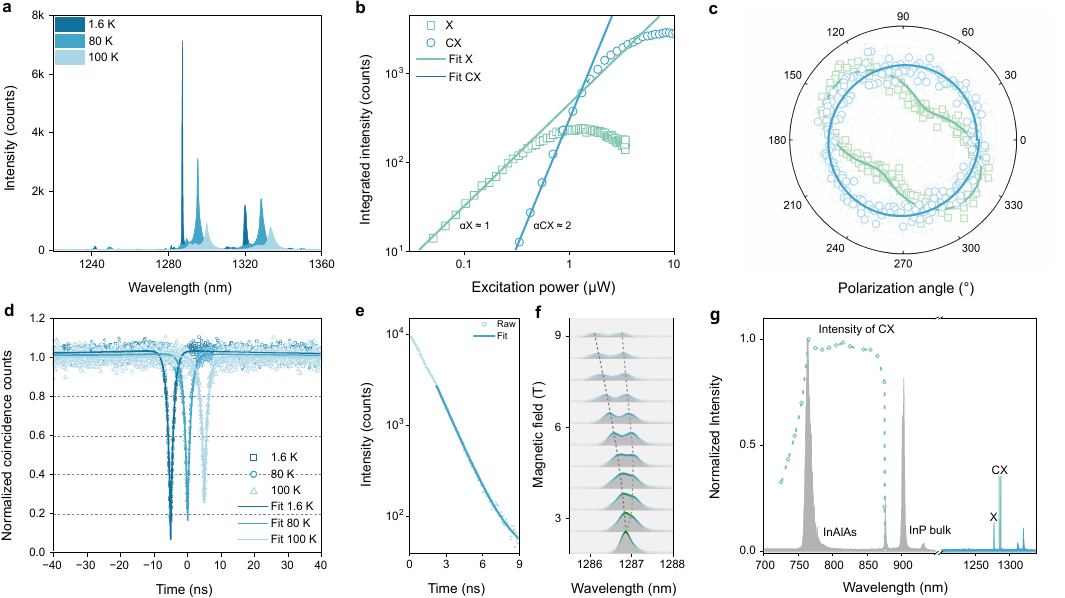}
\caption{\textbf{Telecom single-photon emission from a spectrally isolated emitter.}
\textbf{a} Temperature-dependent $\mu$PL spectra of QD2 in a truncated-cone structure fabricated from a capped S9-like sample, acquired under continuous-wave 815~nm excitation at an incident power of 10~$\mu$W.
\textbf{b} Excitation-power dependence of the neutral-exciton (X) and charged-exciton (CX) lines, shown as integrated intensities with power-law fits.
\textbf{c} Linear-polarization dependence of the X and CX transition energies, with sinusoidal fits.
\textbf{d} Second-order autocorrelation $g^{(2)}(\tau)$ of the CX line at 1.6, 80 and 100~K under continuous-wave 815~nm excitation. No background correction was applied; traces are horizontally offset for clarity.
\textbf{e} Time-resolved decay of the CX line under pulsed 815~nm excitation, with an empirical biexponential fit.
\textbf{f} Magnetic-field-dependent $\mu$PL spectra of QD2 in Faraday geometry; dashed guides trace the Zeeman-split CX branches.
\textbf{g} Photoluminescence-excitation scan of the integrated QD2 CX signal acquired at an incident excitation power of approximately 2~$\mu$W, together with representative PL spectra showing the InAlAs-barrier and InP-substrate emission bands and the monitored X/CX lines.}
\label{Figure2}
\end{figure*}

The corresponding capped samples show different low-temperature $\mu$PL spectra (Fig.~\ref{Figure1}c). S5 shows weak telecom-range emission, whereas S7 shows little telecom emission apart from occasional residual peaks. S9 shows strong telecom emission. The island-forming path therefore yields the strongest telecom emission among these controls. Samples prepared under these island-forming conditions were used for the emitter measurements in Fig.~\ref{Figure2}.

\subsection{Telecom single-photon emission from isolated emitters}
Arrays of truncated-cone structures were fabricated on capped S9-like samples to enhance light extraction for $\mu$PL measurements of individual telecom emitters (Methods). QD2 shows two lines near 1278 and 1287~nm that remain spectrally resolved from 1.6 to 100~K (Fig.~\ref{Figure2}a).

Excitation-power and polarization measurements support the assignment of the 1278 and 1287~nm lines to the neutral exciton (X) and charged exciton (CX), respectively (Fig.~\ref{Figure2}b, c). Over the fitted low-power range, the X intensity is approximately linear in excitation power, whereas the CX intensity is superlinear. The X line also shows a stronger apparent polarization modulation than the CX line.

Second-order autocorrelation measurements establish single-photon emission from the CX line. The fitted $g^{(2)}(0)$ is 0.07 at 1.6~K and remains below 0.25 at 80 and 100~K (Fig.~\ref{Figure2}d). Under pulsed excitation, the CX decay is described empirically by a biexponential function, giving an effective lifetime of about 1.6~ns at 1.6~K (Fig.~\ref{Figure2}e). Two additional emitters give fitted $g^{(2)}(0)$ values of 0.16 and 0.073 (Supplementary Fig.~3).

In Faraday geometry, the CX line shows resolvable Zeeman splitting (Fig.~\ref{Figure2}f), with a linear fit giving an effective optical $g$ factor of $|g_{\mathrm{eff}}|=0.88\pm0.03$. In photoluminescence-excitation (PLE) measurements, the integrated QD2 CX signal peaks near 770~nm in the InAlAs-related spectral region and decreases sharply toward the InP-related region around 875~nm (Fig.~\ref{Figure2}g). The X and CX lines remain spectrally resolved throughout the scan. Additional optical fitting and analysis details are provided in Supplementary Note~4.

\begin{figure*}[!htbp]
\centering
\includegraphics[width=0.98\linewidth]{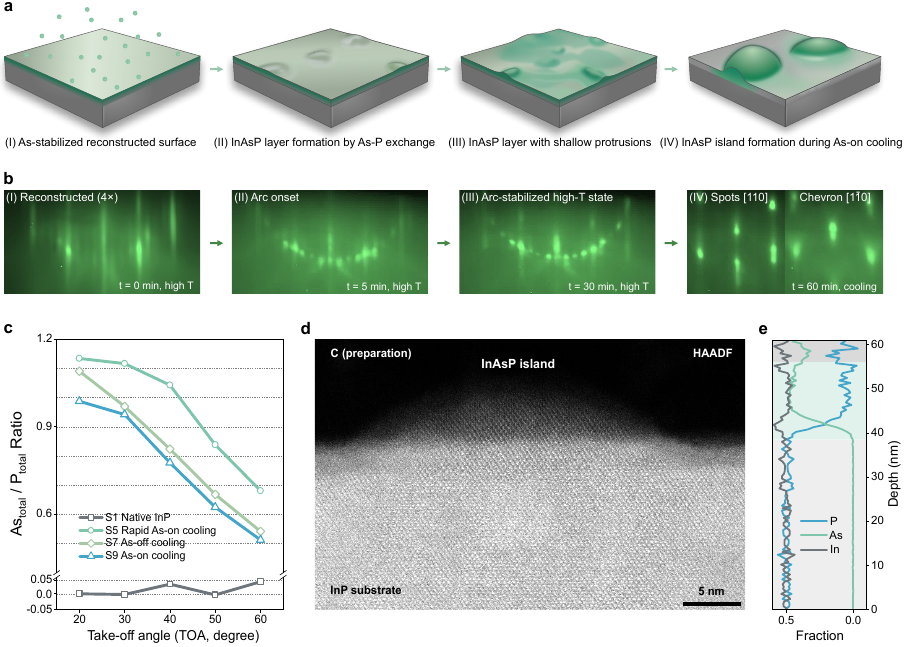}
\caption{\textbf{Surface evolution, near-surface As enrichment and InAsP island structure.}
\textbf{a} Schematic four-stage sequence along the S9 As-on path, based on the combined controls and measurements: an As-stabilized reconstructed surface (I), InAsP layer formation by As--P exchange (II), an InAsP layer with shallow protrusions (III), and InAsP island formation during As-on cooling (IV).
\textbf{b} Representative reflection high-energy electron diffraction (RHEED) frames acquired near the corresponding stages in \textbf{a}, showing the $4\times$ reconstructed surface, arc onset, the arc-stabilized high-temperature pattern and spotty/chevron-like patterns during cooling.
\textbf{c} Angle-resolved X-ray photoelectron spectroscopy (ARXPS), shown as \(\mathrm{As}_{\mathrm{total}}/\mathrm{P}_{\mathrm{total}}\) versus take-off angle for S1, S5, S7 and S9. Oxide-separated ratios and oxide fractions are shown in Supplementary Fig.~4; angle-dependent spectra and representative constrained fits are provided in Supplementary Note~5.
\textbf{d} Representative cross-sectional HAADF-STEM image of an island in a lamella prepared by focused ion beam (FIB) milling from the S9 sample.
\textbf{e} Representative normalized In, As and P EELS depth profile acquired from a separate FIB-prepared lamella of the same S9 sample, traversing an island and the surrounding InP region. The carbon protection layer is not part of the epitaxial structure. The corresponding STEM image used for the EELS acquisition, elemental maps and raw line-scan data are shown in Supplementary Fig.~5.}
\label{Figure3}
\end{figure*}

\subsection{Surface evolution and composition of the InAsP islands}
Figure~\ref{Figure3}a summarizes four stages along the CPSA route. Representative RHEED frames show the corresponding \textit{in situ} evolution (Fig.~\ref{Figure3}b). RHEED first exhibits a streaky $4\times$ pattern, consistent with surface structures reported previously for As-exposed InP(100) \cite{averbeck_oxide_1991,Li2002AsExchangeInP,kolosovsky_determination_2022}. With continued As exposure, weak arc-like features appear and later change little, marking arc onset and the arc-stabilized stage. Similar curved Laue-zone envelopes formed by diffraction spots have been reported for several reconstructed GaAs(001) surfaces \cite{ohtake_surface_2008}. The arc-like pattern alone therefore does not identify large three-dimensional islands. During cooling, the pattern becomes spotty and chevron-like, resembling RHEED signatures reported during three-dimensional InAs islanding and faceting \cite{hanada_anisotropic_2001,Gutierrez2001FacetingAPL}. Together with the growth-interruption controls, these observations place large three-dimensional island formation after the arc-stabilized high-temperature stage.

For S5, S7 and S9, \(\mathrm{As}_{\mathrm{total}}/\mathrm{P}_{\mathrm{total}}\) decreases with increasing take-off angle (Fig.~\ref{Figure3}c). This angular dependence indicates As enrichment near the surface that decreases toward the InP substrate. Because oxide contributions differ among samples, the absolute ratios cannot be compared directly across the sample series. The total As signal includes both In-bonded and oxidized As. We therefore use the angular dependence to identify the near-surface gradient, rather than the absolute ratios to quantify retained As. Oxide-separated ratios and oxide fractions are shown in Supplementary Fig.~4, with representative spectra and fitting details provided in Supplementary Note~5.

A representative cross-sectional HAADF-STEM image of S9 shows continuous lattice planes across the island--substrate interface, with no obvious extended defects within the field of view (Fig.~\ref{Figure3}d). A normalized In, As and P EELS depth profile from a separate lamella prepared by focused ion beam milling shows an As-rich, P-depleted island region and a compositionally graded transition into the surrounding InP matrix (Fig.~\ref{Figure3}e). Together, the composition measurements support formation of an As-rich near-surface InAsP layer by high-temperature As--P exchange and identify the final islands as graded, As-rich InAsP. The growth-interruption experiments place the formation of the large three-dimensional islands during As-on cooling.

\begin{figure*}[!htbp]
\centering
\includegraphics[width=0.98\linewidth]{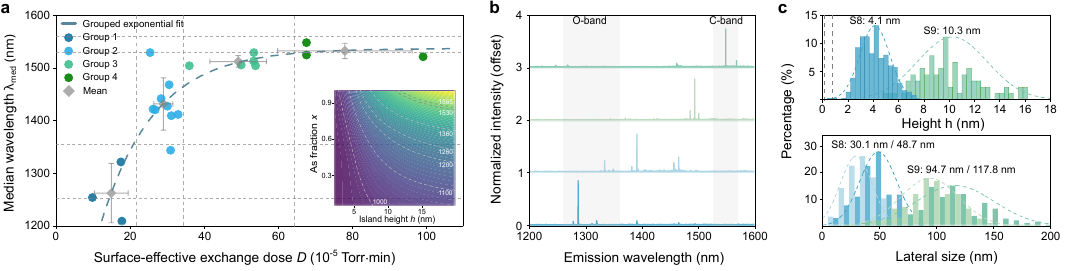}
\caption{\textbf{Surface-effective exchange-dose calibration and island morphology.}
\textbf{a} Median emission wavelength $\lambda_{\mathrm{med}}$ of 20 wafers grown with As-on cooling as a function of the surface-effective exchange dose $D$, reported in units of $10^{-5}~\mathrm{Torr}\cdot\mathrm{min}$. Small solid circles show individual wafer medians; larger solid diamonds show mean values within four adjacent $D$ ranges. Horizontal and vertical error bars indicate the corresponding within-group standard deviations in $D$ and $\lambda_{\mathrm{med}}$, respectively. Vertical dashed lines separate the four $D$ groups, and horizontal dashed lines mark the O- and C-band wavelength boundaries. The dashed curve is a saturating exponential fit to the grouped means. Inset: calculated ground-state transition wavelength for lens-shaped \(\mathrm{InAs}_{\mathrm{x}}\mathrm{P}_{\mathrm{1-x}}\) islands embedded in InP as a function of island height $h$ and As fraction $x$; dashed contours and labels denote equal transition wavelengths.
\textbf{b} Normalized 4~K $\mu$PL spectra from one capped sample in each of the four $D$ groups, collectively spanning 1.2--1.6~\textmu m. Spectra are vertically offset for clarity; gray shading marks the O- and C-band wavelength windows.
\textbf{c} AFM-derived island statistics for S8 and S9. The upper panel shows height distributions, and the lower panel shows lateral-size distributions measured along the minor and major island axes. Histograms are shown as percentages and were extracted from a central $5\times5~\mu$m$^2$ region of each sample ($N=626$ for S8 and $N=187$ for S9); dashed curves are Gaussian fits to the corresponding distributions, and the labeled values are the fitted Gaussian centers. Dashed reference lines in the height histogram indicate the characteristic vertical scales of the reference InP surface (S1; $<0.2$~nm) and the S5 surface with shallow protrusions ($<1$~nm).}
\label{Figure4}
\end{figure*}

\subsection{Wafer-level wavelength calibration}
Across 20 wafers grown under As-on cooling conditions, the median emission wavelength $\lambda_{\mathrm{med}}$ shifts to longer wavelengths with increasing surface-effective exchange dose $D$ (Fig.~\ref{Figure4}a). To compare wafers grown with different As fluxes and temperature--time paths, we define
\begin{equation}
D = \Phi_{\mathrm{As}}\left[t_{\mathrm{high}} + w_{\mathrm{cool}}\left(t_{\mathrm{cool}}+t_{\mathrm{low}}\right)\right],
\end{equation}
where $\Phi_{\mathrm{As}}$ is the applied As flux, recorded as the beam-equivalent pressure (BEP), and $t_{\mathrm{high}}$, $t_{\mathrm{cool}}$ and $t_{\mathrm{low}}$ are the logged durations of the high-temperature, cooling and low-temperature windows, respectively. The high-temperature window starts when the $4\times$ pattern first appears and ends when cooling begins. It includes arc onset and the arc-stabilized stage.

We calculate the lower-temperature weighting factor as
\begin{equation}
w_{\mathrm{cool}}=
\operatorname{clip}\left(
\frac{T_{\mathrm{low}}-380^{\circ}\mathrm{C}}
{T_{\mathrm{high}}-380^{\circ}\mathrm{C}},
0,1
\right),
\end{equation}
where $T_{\mathrm{high}}$ and $T_{\mathrm{low}}$ are the logged high- and low-temperature bounds. Here, $\operatorname{clip}(y,0,1)$ restricts $y$ to the interval from 0 to 1. The lower-temperature weighting represents the reported weakening of substitutional As--P exchange as temperature decreases on InP(100) \cite{Li2002AsExchangeInP,Berdnikov2024}. We use $380^{\circ}$C as a reference temperature rather than a sharp physical threshold, and fix the weighting from the logged temperatures rather than the wavelength data. Spotty and chevron-like RHEED features nevertheless emerge during cooling, showing that island development can continue as substitutional exchange becomes less effective. Alternative reference temperatures and weighting schemes change the numerical $D$ values but preserve the wafer ordering and monotonic dose--wavelength trend (Supplementary Note~3). We therefore use $D$ to compare effective As exposure for the As-on growth conditions studied here, not as a direct measure of retained As.

The individual wafer medians show some scatter, but the mean values for four adjacent $D$ ranges shift monotonically to longer wavelengths. The red-shift becomes smaller at larger $D$. We describe this trend by
\begin{equation}
\lambda(D)=\lambda_{\mathrm{sat}}-\Delta\lambda\exp\left(-\frac{D}{D_0}\right).
\end{equation}
The fit gives $\lambda_{\mathrm{sat}}\approx1538$~nm, $\Delta\lambda\approx738$~nm and $D_0\approx15.1\times10^{-5}~\mathrm{Torr}\cdot\mathrm{min}$. We use this relation as an empirical wafer-level calibration for the As-on growth conditions studied here. Representative spectra from the four $D$ groups collectively span 1.2--1.6~\textmu m (Fig.~\ref{Figure4}b). The pooled emitter distributions retain the same group ordering and show both within-wafer and wafer-to-wafer spread (Supplementary Fig.~6).

To compare island morphology within the As-on-cooled samples, we use S8 and S9 as representative lower- and higher-$D$ conditions, respectively (Fig.~\ref{Figure4}c). S8 contains smaller islands with a local areal density of approximately $25~\mu\mathrm{m}^{-2}$, whereas S9 contains larger islands with a lower density of approximately $7.5~\mu\mathrm{m}^{-2}$. Additional AFM maps and height--lateral-size comparisons are provided in Supplementary Fig.~2. The change from smaller, denser islands in S8 to larger, sparser islands in S9 accompanies the increase in $D$, but does not establish a one-to-one relation between island dimensions and individual emitter wavelength.

The effective-mass calculation in Fig.~\ref{Figure4}a tests whether the observed red-shift is consistent with changes in island size and composition. For lens-shaped \(\mathrm{InAs}_{\mathrm{x}}\mathrm{P}_{\mathrm{1-x}}\) islands embedded in InP, increasing either island height $h$ or As fraction $x$ shifts the calculated ground-state transition to longer wavelengths. The calculated size trend is consistent with the larger islands measured under the higher-$D$ S9 condition. STEM/EELS measurements identify the islands as As-rich InAsP (Fig.~\ref{Figure3}), so variations in $x$ could contribute to the wavelength shift in the same direction. Because size and composition can change together, the calculation is not used to separate their contributions or infer an As fraction from the optical data.

\section{Discussion}

\begin{table*}[t]

\makeatletter
\long\def\@makecaption#1#2{%
  \par
  \vskip\abovecaptionskip
  \begingroup
    \small\rmfamily
    \noindent\RaggedRight
    #1\enspace #2\par
  \endgroup
  \vskip\belowcaptionskip
}
\makeatother

\caption{\textbf{\textbar\ Epitaxial routes and material systems for telecom-band quantum-dot emitters.}}
\label{tab:qd_platforms_main}

\small
\setlength{\tabcolsep}{2.5pt}
\renewcommand{\arraystretch}{1.15}

\noindent

\begin{tabular}{@{}lllll@{}}
\hline

\parbox[t]{0.21\textwidth}{\raggedright \textbf{Route}} &
\parbox[t]{0.15\textwidth}{\raggedright \textbf{Material system}} &
\parbox[t]{0.12\textwidth}{\raggedright \textbf{Emission band}} &
\parbox[t]{0.37\textwidth}{\raggedright \textbf{Defining growth sequence}} &
\parbox[t]{0.11\textwidth}{\raggedright \textbf{References}}\tabularnewline
\hline

\parbox[t]{0.21\textwidth}{\raggedright Stranski--Krastanov (SK)} &
\parbox[t]{0.15\textwidth}{\raggedright InAs/InGaAs/GaAs} &
\parbox[t]{0.12\textwidth}{\raggedright O/C} &
\parbox[t]{0.37\textwidth}{\raggedright InAs-based deposition $\rightarrow$ strain-driven islanding.} &
\parbox[t]{0.11\textwidth}{\raggedright \cite{Kolatschek2021BrightPurcellOband,Nawrath2023BrightCband}}\tabularnewline
\hline

\parbox[t]{0.21\textwidth}{\raggedright Stranski--Krastanov (SK)} &
\parbox[t]{0.15\textwidth}{\raggedright InAs/InP; InAs/InAlGaAs/InP} &
\parbox[t]{0.12\textwidth}{\raggedright C} &
\parbox[t]{0.37\textwidth}{\raggedright InAs deposition $\rightarrow$ strain-driven islanding.} &
\parbox[t]{0.11\textwidth}{\raggedright \cite{miyazawa_single-photon_2016,Holewa2024TelecomSingleEmitter,Hauser2026TelecomIndistinguishable}}\tabularnewline
\hline

\parbox[t]{0.21\textwidth}{\raggedright Droplet epitaxy (DE)} &
\parbox[t]{0.15\textwidth}{\raggedright InAs/InP} &
\parbox[t]{0.12\textwidth}{\raggedright C} &
\parbox[t]{0.37\textwidth}{\raggedright Group-III droplet formation $\rightarrow$ crystallization under group-V flux.} &
\parbox[t]{0.11\textwidth}{\raggedright \cite{Gurioli2019DropletReview,Holewa2022DropletInAsInP}}\tabularnewline
\hline

\parbox[t]{0.21\textwidth}{\raggedright Local droplet etching (LDE)} &
\parbox[t]{0.15\textwidth}{\raggedright InGaAs/InAlAs/InP; GaSb/AlGaSb} &
\parbox[t]{0.12\textwidth}{\raggedright C/S} &
\parbox[t]{0.37\textwidth}{\raggedright Group-III droplet formation $\rightarrow$ nanohole etching $\rightarrow$ semiconductor infill.} &
\parbox[t]{0.11\textwidth}{\raggedright \cite{Polczynska2026LDETelecomCband,Michl2023StrainFreeGaSb,Masson2026NanoholeTelecom}}\tabularnewline
\hline

\parbox[t]{0.21\textwidth}{\raggedright \textbf{Cooling-path-selected self-assembly (CPSA)}} &
\parbox[t]{0.15\textwidth}{\raggedright \textbf{InAsP/InP(001)}} &
\parbox[t]{0.12\textwidth}{\raggedright \textbf{O/E/S/C/L}} &
\parbox[t]{0.37\textwidth}{\raggedright \textbf{High-temperature As--P exchange $\rightarrow$ As-on cooling $\rightarrow$ InAsP island formation.}} &
\parbox[t]{0.11\textwidth}{\raggedright \textbf{This work}}\tabularnewline
\hline

\end{tabular}

\end{table*}

These results establish a cooling-path-selected route to telecom InAsP QDs on InP(001). High-temperature As exposure first forms an As-enriched near-surface InAsP layer. The subsequent cooling path determines whether this layer develops into large three-dimensional islands. Table~\ref{tab:qd_platforms_main} compares CPSA with representative telecom QD routes according to the step that drives island formation. In SK growth, deposition of lattice-mismatched dot material leads to strain-driven islanding \cite{Tersoff1993,Shchukin1999Spontaneous}. Droplet-based routes instead use group-III droplets either as the QD precursor or to create nanohole templates for later infill \cite{Gurioli2019DropletReview,Holewa2022DropletInAsInP,Polczynska2026LDETelecomCband}. CPSA differs in that islands form during cooling from a near-surface InAsP layer generated by high-temperature As--P exchange. Earlier studies separately reported As--P-exchanged layers, continued exchange during cooling, As-rich islands and cooling-associated ripening of intentionally deposited InAs \cite{Taskinen1997InAsInPIslands,Li2002AsExchangeInP,leavitt_pathway_2015,Wang1998AsPExchangeInAsQD,kolosovsky_determination_2022,Yacob2014Ripening,Kors2018Ripening}. However, they did not establish how temperature--time and As-on/As-off cooling paths select the final morphology of an exchange-generated near-surface InAsP layer. The present control series establishes this relation and defines a reproducible route to InAsP islands without intentional deposition of group-III dot material. The closest comparison is LDE, because both routes use temperature and group-V conditions to control surface mass transport. LDE interrupts the group-V flux, deposits group-III droplets to etch recessed nanoholes and then fills them with QD material \cite{Polczynska2026LDETelecomCband,Michl2023StrainFreeGaSb,Masson2026NanoholeTelecom}. CPSA instead maintains As exposure during cooling, allowing an exchange-modified near-surface InAsP layer to develop into protruding islands without separate droplet formation, etching or infill. The two routes therefore form a mirror-image contrast in sequence and geometry, although this analogy does not imply a microscopic reversal of the LDE mechanism.

The control series shows that the cooling path, including the temperature--time trajectory and As-flux condition, selects the final morphology. High-temperature As treatment prepares the near-surface InAsP layer but does not determine whether large islands form. Within the tested paths, slow As-on cooling followed by a low-temperature hold produces large islands. Rapid cooling retains a nearly flat surface or only shallow protrusions, whereas As-off cooling produces no large islands. We do not assign the shallow protrusions to a specific microscopic structure or use them to determine when nucleation occurred. Because AFM records the surface only after cooling, the rapid-cooling samples represent path-interruption endpoints rather than instantaneous snapshots, and the stage of large-island development is therefore identified from the growth-interruption and As-on/As-off cooling controls together with the \textit{in situ} RHEED sequence. The arc-stabilized pattern is reached at high temperature and does not by itself identify large islands or show that surface evolution has ceased; the combined evidence instead places large-island development after this stage and along the slow As-on cooling path. Spotty and chevron-like features emerge later along the island-forming path, consistent with three-dimensional roughening and faceting, respectively. The same arc-stabilized state can therefore lead to different final morphologies, showing that the cooling path, rather than the high-temperature treatment alone, determines the result. Combined with ARXPS and STEM/EELS, these controls support retention and redistribution of an As-rich near-surface InAsP layer into graded, coherently embedded islands during cooling under As.

A qualitative Gibbs free-energy balance helps explain this path dependence. As--P exchange changes local composition and lattice parameter. The relevant balance therefore includes exchange-related chemical, surface/interface and elastic contributions. Cooling changes the relative free energies of the accessible surface states. It also changes the rates of As--P exchange, surface diffusion and group-V desorption. The As flux most directly changes the As--P chemical potential and near-surface As retention. The RHEED sequence evolves over several minutes: arc-like features develop at high temperature, whereas spotty and chevron-like features appear only after several minutes at lower temperature. Slow As-on cooling followed by the low-temperature hold therefore provides more time for the near-surface InAsP material to redistribute into islands.

Rapid cooling shortens this interval, whereas closing the As shutter stops the As supply and reduces near-surface As retention. For the representative island geometries, the estimates in Supplementary Note~2 indicate that the characteristic magnitude of the exchange-related chemical contribution exceeds those of the corresponding surface/interface and coherent elastic contributions. The As-on/As-off condition therefore most directly affects this chemical contribution. Surface/interface and elastic contributions nevertheless remain important in determining island shape and strain accommodation. Lattice continuity across the island--substrate interface and the localized strain measured by geometric phase analysis (GPA) show that elastic effects remain present in the final structure. However, the controls do not identify strain alone as the origin of the path-dependent outcome. The estimates support this thermodynamic picture but compare only characteristic energy scales; they do not define a kinetic model or a unique atomistic pathway.

Beyond determining whether islands form, CPSA provides a wafer-level calibration of the emission wavelength. Across 20 wafers grown with As-on cooling, the median wavelength shifts monotonically to longer wavelengths with increasing surface-effective exchange dose $D$. The corresponding emitter populations span 1.2--1.6~\textmu m. The dose $D$ combines the high-temperature As exposure with the effective contribution from As-on cooling into a single empirical growth variable. At larger $D$, further increases in dose produce smaller wavelength shifts, consistent with a reduced effect of additional As exposure as exchange progresses. Because the emission wavelength depends on both island size and composition, we use the fitted exponential relation as a practical growth calibration for the tested conditions rather than a kinetic law.

More broadly, controlled cooling can therefore serve as an active epitaxial step after high-temperature surface exchange. Whether the same principle applies to other III--V systems in which group-V exchange occurs remains to be tested \cite{Losurdo2006AnionExchange}. On InP, the next practical step is to combine CPSA with charge stabilization and nanophotonic structures to develop electrically controlled, integrated telecom quantum-light sources.

\bibliographystyle{naturemag}
\bibliography{reference}

@article{Somaschi2016NearOptimal,
  author        = {Somaschi, N. and Giesz, V. and De Santis, L. and Loredo, J. C. and Almeida, M. P. and Hornecker, G. and Portalupi, S. L. and Grange, T. and Ant{\'o}n, C. and Demory, J. and G{\'o}mez, C. and Sagnes, I. and Lanzillotti-Kimura, N. D. and Lema{\^i}tre, A. and Auff{\`e}ves, A. and White, A. G. and Lanco, L. and Senellart, P.},
  title         = {Near-optimal single-photon sources in the solid state},
  journal       = {Nat. Photonics},
  volume        = {10},
  pages         = {340--345},
  year          = {2016},
  doi           = {10.1038/nphoton.2016.23}
}

@article{Wang2019OptimalSPS,
  author        = {Wang, Hui and He, Yu-Ming and Chung, Tung Hsun and Hu, Hai and Yu, Ying and Chen, Si and Ding, Xing and Chen, Ming-Cheng and Qin, Jian and Yang, Xiaoxia and Liu, Run-Ze and Duan, Zhao-Chen and Li, Jin-Peng and Gerhardt, Stefan and Winkler, Karol and Jurkat, Jonathan and Wang, Lin-Jun and Gregersen, Niels and Huo, Yong-Heng and Dai, Qing and Yu, Siyuan and H{\"o}fling, Sven and Lu, Chao-Yang and Pan, Jian-Wei},
  title         = {Towards optimal single-photon sources from polarized microcavities},
  journal       = {Nat. Photonics},
  volume        = {13},
  pages         = {770--775},
  year          = {2019},
  doi           = {10.1038/s41566-019-0494-3}
}

@article{Liu2019EntangledPhotonPairs,
  author        = {Liu, J. and Su, R. and Wei, Y. and Yao, B. and da Silva, S. F. C. and Yu, Y. and Iles-Smith, J. and Srinivasan, K. and Rastelli, A. and Li, J. and Wang, X.},
  title         = {A solid-state source of strongly entangled photon pairs with high brightness and indistinguishability},
  journal       = {Nat. Nanotechnol.},
  volume        = {14},
  pages         = {586--593},
  year          = {2019},
  doi           = {10.1038/s41565-019-0435-9}
}

@article{Cogan2023ClusterState,
  author        = {Cogan, Dan and Su, Zu-En and Kenneth, Oded and Gershoni, David},
  title         = {Deterministic generation of indistinguishable photons in a cluster state},
  journal       = {Nat. Photonics},
  volume        = {17},
  pages         = {324--329},
  year          = {2023},
  doi           = {10.1038/s41566-022-01152-2}
}

@article{Wang2020IntegratedPhotonicTechnologies,
  author        = {Wang, Jianwei and Sciarrino, Fabio and Laing, Anthony and Thompson, Mark G.},
  title         = {Integrated photonic quantum technologies},
  journal       = {Nat. Photonics},
  volume        = {14},
  pages         = {273--284},
  year          = {2020},
  doi           = {10.1038/s41566-019-0532-1}
}

@article{Maring2024SinglePhotonQC,
  author        = {Maring, Nicolas and Fyrillas, Andreas and Pont, Mathias and Ivanov, Edouard and Stepanov, Petr and Margaria, Nico and Hease, William and Pishchagin, Anton and Lema{\^i}tre, Aristide and Sagnes, Isabelle and Au, Thi Huong and Boissier, S{\'e}bastien and Bertasi, Eric and Baert, Aur{\'e}lien and Valdivia, Mario and Billard, Marie and Acar, Ozan and Brieussel, Alexandre and Mezher, Rawad and Wein, Stephen C. and Salavrakos, Alexia and Sinnott, Patrick and Fioretto, Dario A. and Emeriau, Pierre-Emmanuel and Belabas, Nadia and Mansfield, Shane and Senellart, Pascale and Senellart, Jean and Somaschi, Niccolo},
  title         = {A versatile single-photon-based quantum computing platform},
  journal       = {Nat. Photonics},
  volume        = {18},
  pages         = {603--609},
  year          = {2024},
  doi           = {10.1038/s41566-024-01403-4}
}

@article{Wang2025ScalablePhotonic,
  author        = {Wang, Hui and Ralph, Timothy C. and Renema, Jelmer J. and Lu, Chao-Yang and Pan, Jian-Wei},
  title         = {Scalable photonic quantum technologies},
  journal       = {Nat. Mater.},
  volume        = {24},
  number        = {12},
  pages         = {1883--1897},
  year          = {2025},
  doi           = {10.1038/s41563-025-02306-7}
}

@article{Yu2023TelecomReview,
  author        = {Yu, Ying and Liu, Shunfa and Lee, Chang-Min and Michler, Peter and Reitzenstein, Stephan and Srinivasan, Kartik and Waks, Edo and Liu, Jin},
  title         = {Telecom-band quantum dot technologies for long-distance quantum networks},
  journal       = {Nat. Nanotechnol.},
  volume        = {18},
  pages         = {1389--1400},
  year          = {2023},
  doi           = {10.1038/s41565-023-01528-7}
}

@article{Hauser2026TelecomIndistinguishable,
  author        = {Hauser, Nico and Bayerbach, Matthias and Kaupp, Jochen and others},
  title         = {Deterministic and highly indistinguishable single photons in the telecom {C}-band},
  journal       = {Nat. Commun.},
  volume        = {17},
  pages         = {537},
  year          = {2026},
  doi           = {10.1038/s41467-026-68336-0}
}

@article{Holewa2024TelecomSingleEmitter,
  author        = {Holewa, Pawe{\l} and Vajner, Daniel A. and Zi{\k{e}}ba-Ost{\'o}j, Emilia and others},
  title         = {High-throughput quantum photonic devices emitting indistinguishable photons in the telecom {C}-band},
  journal       = {Nat. Commun.},
  volume        = {15},
  pages         = {3358},
  year          = {2024},
  doi           = {10.1038/s41467-024-47551-7}
}

@article{Muller2018TelecomQLED,
  author        = {M{\"u}ller, T. and Skiba-Szymanska, J. and Krysa, A. B. and Huwer, J. and Felle, M. and Anderson, M. and Stevenson, R. M. and Heffernan, J. and Ritchie, D. A. and Shields, A. J.},
  title         = {A quantum light-emitting diode for the standard telecom window around 1,550 nm},
  journal       = {Nat. Commun.},
  volume        = {9},
  pages         = {862},
  year          = {2018},
  doi           = {10.1038/s41467-018-03251-7}
}

@article{Anderson2020QuantumTeleportation,
  author        = {Anderson, M. and M{\"u}ller, T. and Huwer, J. and Skiba-Szymanska, J. and Krysa, A. B. and Stevenson, R. M. and Heffernan, J. and Ritchie, D. A. and Shields, A. J.},
  title         = {Quantum teleportation using highly coherent emission from telecom {C}-band quantum dots},
  journal       = {npj Quantum Inf.},
  volume        = {6},
  pages         = {14},
  year          = {2020},
  doi           = {10.1038/s41534-020-0249-5}
}

@article{Laccotripes2024SpinPhotonTelecom,
  author        = {Laccotripes, P. and M{\"u}ller, T. and Stevenson, R. M. and Skiba-Szymanska, J. and Ritchie, D. A. and Shields, A. J.},
  title         = {Spin-photon entanglement with direct photon emission in the telecom {C}-band},
  journal       = {Nat. Commun.},
  volume        = {15},
  pages         = {9740},
  year          = {2024},
  doi           = {10.1038/s41467-024-53964-1}
}

@article{Yang2024IntercityQKD,
  author        = {Yang, Jingzhong and Jiang, Zenghui and Benthin, Frederik and Hanel, Joscha and Fandrich, Tom and Joos, Raphael and Bauer, Stephanie and Kolatschek, Sascha and Hreibi, Ali and Rugeramigabo, Eddy Patrick and Jetter, Michael and Portalupi, Simone Luca and Zopf, Michael and Michler, Peter and K{\"u}ck, Stefan and Ding, Fei},
  title         = {High-rate intercity quantum key distribution with a semiconductor single-photon source},
  journal       = {Light Sci. Appl.},
  volume        = {13},
  pages         = {150},
  year          = {2024},
  doi           = {10.1038/s41377-024-01488-0}
}

@article{Kolatschek2021BrightPurcellOband,
  author        = {Kolatschek, Sascha and Nawrath, Cornelius and Bauer, Stephanie and Huang, Jiasheng and Fischer, Julius and Sittig, Robert and Jetter, Michael and Portalupi, Simone Luca and Michler, Peter},
  title         = {Bright {Purcell} Enhanced Single-Photon Source in the Telecom {O}-band Based on a Quantum Dot in a Circular {Bragg} Grating},
  journal       = {Nano Lett.},
  volume        = {21},
  number        = {18},
  pages         = {7740--7745},
  year          = {2021},
  doi           = {10.1021/acs.nanolett.1c02647}
}

@article{Nawrath2023BrightCband,
  author        = {Nawrath, Cornelius and Joos, Raphael and Kolatschek, Sascha and Bauer, Stephanie and Pruy, Pascal and Hornung, Florian and Fischer, Julius and Huang, Jiasheng and Vijayan, Ponraj and Sittig, Robert and Jetter, Michael and Portalupi, Simone Luca and Michler, Peter},
  title         = {Bright Source of {Purcell}-Enhanced, Triggered, Single Photons in the Telecom {C}-band},
  journal       = {Adv. Quantum Technol.},
  volume        = {6},
  number        = {11},
  pages         = {2300111},
  year          = {2023},
  doi           = {10.1002/qute.202300111}
}

@article{Michl2023StrainFreeGaSb,
  author        = {Michl, Johannes and Peniakov, Giora and Pfenning, Andreas and Hilska, Joonas and Chellu, Abhiroop and Bader, Andreas and Guina, Mircea and H{\"o}fling, Sven and Hakkarainen, Teemu and Huber-Loyola, Tobias},
  title         = {Strain-Free {GaSb} Quantum Dots as Single-Photon Sources in the Telecom {S}-band},
  journal       = {Adv. Quantum Technol.},
  volume        = {6},
  pages         = {2300180},
  year          = {2023},
  doi           = {10.1002/qute.202300180}
}

@article{miyazawa_single-photon_2016,
  author        = {Miyazawa, T. and Takemoto, K. and Nambu, Y. and Miki, S. and Yamashita, T. and Terai, H. and Fujiwara, M. and Sasaki, M. and Sakuma, Y. and Takatsu, M. and Yamamoto, T. and Arakawa, Y.},
  title         = {Single-photon emission at 1.5~{$\mu$m} from an {InAs/InP} quantum dot with highly suppressed multi-photon emission probabilities},
  journal       = {Appl. Phys. Lett.},
  volume        = {109},
  number        = {13},
  pages         = {132106},
  year          = {2016},
  doi           = {10.1063/1.4961888}
}

@article{Holewa2022DropletInAsInP,
  author        = {Holewa, Pawe{\l} and Kadkhodazadeh, Shima and Gawe{\l}czyk, Micha{\l} and Ba{\l}uta, Pawe{\l} and Musia{\l}, Anna and Dubrovskii, Vladimir G. and Syperek, Marcin and Semenova, Elizaveta},
  title         = {Droplet epitaxy symmetric {InAs/InP} quantum dots for quantum emission in the third telecom window: morphology, optical and electronic properties},
  journal       = {Nanophotonics},
  volume        = {11},
  number        = {8},
  pages         = {1515--1526},
  year          = {2022},
  doi           = {10.1515/nanoph-2021-0482}
}

@misc{Polczynska2026LDETelecomCband,
  author        = {Po{\l}czy{\'n}ska, Karolina E. and Wyborski, Pawe{\l} and Gawe{\l}czyk, Micha{\l} and Kadkhodazadeh, Shima and Munkhbat, Battulga and Sanguinetti, Stefano and Semenova, Elizaveta},
  title         = {Local droplet etching-assisted quantum dot epitaxy for telecom {C}-band quantum light emitters},
  year          = {2026},
  doi           = {10.48550/arXiv.2605.05956},
  archiveprefix = {arXiv},
  primaryclass  = {cond-mat.mes-hall},
  note          = {Preprint at arXiv:2605.05956}
}

@article{Gurioli2019DropletReview,
  author        = {Gurioli, Massimo and Wang, Zhiming and Rastelli, Armando and Kuroda, Takashi and Sanguinetti, Stefano},
  title         = {Droplet epitaxy of semiconductor nanostructures for quantum photonic devices},
  journal       = {Nat. Mater.},
  volume        = {18},
  pages         = {799--810},
  year          = {2019},
  doi           = {10.1038/s41563-019-0355-y}
}

@article{Smit2019InPIntegration,
  author        = {Smit, Meint and Williams, Kevin and van der Tol, Jos},
  title         = {Past, present, and future of {InP}-based photonic integration},
  journal       = {APL Photonics},
  volume        = {4},
  number        = {5},
  pages         = {050901},
  year          = {2019},
  doi           = {10.1063/1.5087862}
}

@article{Gendry2004QuantumSticks,
  author        = {Gendry, M. and Monat, C. and Brault, J. and Regreny, P. and Hollinger, G. and Salem, B. and Guillot, G. and Benyattou, T. and Bru-Chevallier, C. and Bremond, G. and Marty, O.},
  title         = {From large to low height dispersion for self-organized {InAs} quantum sticks emitting at 1.55 {$\mu$m} on {InP} (001)},
  journal       = {J. Appl. Phys.},
  volume        = {95},
  number        = {9},
  pages         = {4761--4766},
  year          = {2004},
  doi           = {10.1063/1.1690101}
}

@article{Dery2004QuantumDashNature,
  author        = {Dery, H. and Benisty, E. and Epstein, A. and Alizon, R. and Mikhelashvili, V. and Eisenstein, G. and Schwertberger, R. and Gold, D. and Reithmaier, J. P. and Forchel, A.},
  title         = {On the nature of quantum dash structures},
  journal       = {J. Appl. Phys.},
  volume        = {95},
  number        = {11},
  pages         = {6103--6111},
  year          = {2004},
  doi           = {10.1063/1.1715135}
}

@article{Bierwagen2005VicinalInP,
  author        = {Bierwagen, O. and Masselink, W. T.},
  title         = {Self-organized growth of {InAs} quantum wires and dots on {InP}(001): The role of vicinal substrates},
  journal       = {Appl. Phys. Lett.},
  volume        = {86},
  number        = {11},
  pages         = {113110},
  year          = {2005},
  doi           = {10.1063/1.1884762}
}

@article{Taskinen1997InAsInPIslands,
  author        = {Taskinen, M. and Sopanen, M. and Lipsanen, H. and Tulkki, J. and Tuomi, T. and Ahopelto, J.},
  title         = {Self-organized {InAs} islands on (100) {InP} by metalorganic vapor-phase epitaxy},
  journal       = {Surf. Sci.},
  volume        = {376},
  number        = {1--3},
  pages         = {60--68},
  year          = {1997},
  doi           = {10.1016/S0039-6028(96)01597-X}
}

@article{Yoon1999AsPExchangeInAsInP,
  author        = {Yoon, Sukho and Moon, Youngboo and Lee, Tae-Wan and Yoon, Euijoon and Kim, Young Dong},
  title         = {Effects of {As/P} exchange reaction on the formation of {InAs/InP} quantum dots},
  journal       = {Appl. Phys. Lett.},
  volume        = {74},
  number        = {14},
  pages         = {2029--2031},
  year          = {1999},
  doi           = {10.1063/1.123746}
}

@article{gutierrez_role_2002,
  author        = {Guti{\'e}rrez, H. R. and Cotta, M. A. and Bortoleto, J. R. R. and de Carvalho, M. M. G.},
  title         = {Role of group {V} exchange on the shape and size of {InAs/InP} self-assembled nanostructures},
  journal       = {J. Appl. Phys.},
  volume        = {92},
  number        = {12},
  pages         = {7523--7526},
  year          = {2002},
  doi           = {10.1063/1.1524014}
}

@article{leavitt_pathway_2015,
  author        = {Leavitt, R. P. and Richardson, C. J. K.},
  title         = {Pathway to achieving circular {InAs} quantum dots directly on (100) {InP} and to tuning their emission wavelengths toward 1.55~{$\mu$m}},
  journal       = {J. Vac. Sci. Technol. B},
  volume        = {33},
  number        = {5},
  pages         = {051202},
  year          = {2015},
  doi           = {10.1116/1.4927199}
}

@article{averbeck_oxide_1991,
  author        = {Averbeck, R. and Riechert, H. and Schl{\"o}tterer, H. and Weimann, G.},
  title         = {Oxide desorption from {InP} under stabilizing pressures of {P$_2$} or {As$_4$}},
  journal       = {Appl. Phys. Lett.},
  volume        = {59},
  number        = {14},
  pages         = {1732--1734},
  year          = {1991},
  doi           = {10.1063/1.106233}
}

@article{Li2002AsExchangeInP,
  author        = {Li, C. H. and Li, L. and Law, D. C. and Visbeck, S. B. and Hicks, R. F.},
  title         = {Arsenic adsorption and exchange with phosphorus on indium phosphide (001)},
  journal       = {Phys. Rev. B},
  volume        = {65},
  pages         = {205322},
  year          = {2002},
  doi           = {10.1103/PhysRevB.65.205322}
}

@article{Wang1998AsPExchangeInAsQD,
  author        = {Wang, Benzhong and Zhao, Fanghai and Peng, Yuheng and Jin, Zhi and Li, Yudong and Liu, Shiyong},
  title         = {Self-organized {InAs} quantum dots formation by {As/P} exchange reaction on (001) {InP} substrate},
  journal       = {Appl. Phys. Lett.},
  volume        = {72},
  number        = {19},
  pages         = {2433--2435},
  year          = {1998},
  doi           = {10.1063/1.121396}
}

@article{kolosovsky_determination_2022,
  author        = {Kolosovsky, D. A. and Dmitriev, D. V. and Ponomarev, S. A. and Toropov, A. I. and Zhuravlev, K. S.},
  title         = {Determination of the phosphorus desorption rate during high-temperature annealing of the {InP}(001) substrate in an arsenic flux},
  journal       = {J. Phys. Conf. Ser.},
  volume        = {2227},
  number        = {1},
  pages         = {012004},
  year          = {2022},
  doi           = {10.1088/1742-6596/2227/1/012004}
}

@article{Berdnikov2024,
  author        = {Berdnikov, Yury and Holewa, Pawe{\l} and Kadkhodazadeh, Shima and {\'S}migiel, Jan Miko{\l}aj and Sakanas, Aurimas and Fr{\k{a}}ckowiak, Adrianna and Yvind, Kresten and Syperek, Marcin and Semenova, Elizaveta},
  title         = {Near-critical {Stranski--Krastanov} growth of {InAs/InP} quantum dots},
  journal       = {Sci. Rep.},
  volume        = {14},
  pages         = {23697},
  year          = {2024},
  doi           = {10.1038/s41598-024-70451-1}
}

@article{ohtake_surface_2008,
  author        = {Ohtake, Akihiro},
  title         = {Surface reconstructions on {GaAs}(001)},
  journal       = {Surf. Sci. Rep.},
  volume        = {63},
  number        = {7--8},
  pages         = {295--327},
  year          = {2008},
  doi           = {10.1016/j.surfrep.2008.03.001}
}

@article{hanada_anisotropic_2001,
  author        = {Hanada, Takashi and Koo, Bon-Heun and Totsuka, Hirofumi and Yao, Takafumi},
  title         = {Anisotropic shape of self-assembled {InAs} quantum dots: {Refraction} effect on spot shape of reflection high-energy electron diffraction},
  journal       = {Phys. Rev. B},
  volume        = {64},
  number        = {16},
  pages         = {165307},
  year          = {2001},
  doi           = {10.1103/PhysRevB.64.165307}
}

@article{Gutierrez2001FacetingAPL,
  author        = {Guti{\'e}rrez, H. R. and Cotta, M. A. and de Carvalho, M. M. G.},
  title         = {Faceting evolution during self-assembling of {InAs/InP} quantum wires},
  journal       = {Appl. Phys. Lett.},
  volume        = {79},
  number        = {23},
  pages         = {3854--3856},
  year          = {2001},
  doi           = {10.1063/1.1424476}
}

@article{Tersoff1993,
  author        = {Tersoff, J. and Tromp, R. M.},
  title         = {Shape Transition in Growth of Strained Islands: Spontaneous Formation of Quantum Wires},
  journal       = {Phys. Rev. Lett.},
  volume        = {70},
  number        = {18},
  pages         = {2782--2785},
  year          = {1993},
  doi           = {10.1103/PhysRevLett.70.2782}
}

@article{Shchukin1999Spontaneous,
  author        = {Shchukin, Vitaliy A. and Bimberg, Dieter},
  title         = {Spontaneous ordering of nanostructures on crystal surfaces},
  journal       = {Rev. Mod. Phys.},
  volume        = {71},
  number        = {4},
  pages         = {1125--1171},
  year          = {1999},
  doi           = {10.1103/RevModPhys.71.1125}
}

@article{Losurdo2006AnionExchange,
  author        = {Losurdo, Maria and Capezzuto, Pio and Bruno, Giovanni and Brown, April S. and Brown, Terence and May, Gary},
  title         = {Fundamental reactions controlling anion exchange during mixed anion heterojunction formation: Chemistry of {As}-for-{Sb} and {Sb}-for-{As} exchange reactions},
  journal       = {J. Appl. Phys.},
  volume        = {100},
  number        = {1},
  pages         = {013531},
  year          = {2006},
  doi           = {10.1063/1.2216049}
}

@article{Kozhukhov1997InPAs4,
  author        = {Kozhukhov, A. V. and Revenko, M. A. and Fedorov, A. A. and Konarski, P. and Herman, M. A.},
  title         = {Influence of {Al} adsorption on {In} and {Ga} thermal desorption from {InP} and {GaAs} surfaces heated under {As$_4$} flux},
  journal       = {Thin Solid Films},
  volume        = {306},
  pages         = {248--252},
  year          = {1997},
  doi           = {10.1016/S0040-6090(97)00177-6}
}

@article{Masson2026NanoholeTelecom,
  author        = {Masson, Ian M. and Hageman, Aden and Whittier, Caleb and Montealegre, David and Kamaliya, Bhaveshkumar and Bassim, Nabil D. and Prineas, John P. and Uppu, Ravitej},
  title         = {Engineering Nanohole-Etched Quantum Dots for Telecom-Band Single-Photon Generation},
  journal       = {ACS Nano},
  volume        = {20},
  number        = {3},
  pages         = {2872--2880},
  year          = {2026},
  doi           = {10.1021/acsnano.5c17982}
}

@article{Spitzer2024TelecomOBandLDE,
  author        = {Spitzer, Nikolai and Kersting, Elias and Grell, Meret and Kohminaei, Danial and Schmidt, Marcel and Bart, Nikolai and Wieck, Andreas D. and Ludwig, Arne},
  title         = {Telecom {O}-Band Quantum Dots Fabricated by Droplet Etching},
  journal       = {Crystals},
  volume        = {14},
  number        = {12},
  pages         = {1014},
  year          = {2024},
  doi           = {10.3390/cryst14121014}
}

@article{Albrechtsen2026QuantumCoherent,
  author        = {Albrechtsen, Marcus and Kr{\"u}ger, Severin and Loredo, Juan C. and Stefan, Lucio and Liu, Zhe and Meng, Yu and Niekamp, Lukas L. and Seyschab, Bianca F. and Spitzer, Nikolai and Warburton, Richard J. and Lodahl, Peter and Ludwig, Arne and Midolo, Leonardo},
  title         = {A quantum-coherent photon--emitter interface in the original telecom band},
  journal       = {Nat. Nanotechnol.},
  volume        = {21},
  pages         = {642--647},
  year          = {2026},
  doi           = {10.1038/s41565-026-02156-7}
}

@article{Sasakura2007IndiumFlush,
  author        = {Sasakura, H. and Kayamori, S. and Adachi, S. and Muto, S.},
  title         = {Effect of indium-flush method on the control of photoluminescence energy of highly uniform self-assembled {InAs} quantum dots by slow molecular beam epitaxy growth},
  journal       = {J. Appl. Phys.},
  volume        = {102},
  number        = {1},
  pages         = {013515},
  year          = {2007},
  doi           = {10.1063/1.2752598}
}

@article{Zhu2025HybridQDSiC,
  author        = {Zhu, Yifan and Liu, Runze and Yi, Ailun and others},
  title         = {A hybrid single quantum dot coupled cavity on a {CMOS}-compatible {SiC} photonic chip for {Purcell}-enhanced deterministic single-photon emission},
  journal       = {Light Sci. Appl.},
  volume        = {14},
  pages         = {86},
  year          = {2025},
  doi           = {10.1038/s41377-024-01676-y}
}

@article{Gaur2026ScalableSiteControlledQD,
  author        = {Gaur, Kartik and Barua, Avijit and Tripathi, Sarthak and others},
  title         = {Scalable quantum photonic platform based on site-controlled quantum dots coupled to circular {Bragg} grating resonators},
  journal       = {Light Sci. Appl.},
  volume        = {15},
  pages         = {260},
  year          = {2026},
  doi           = {10.1038/s41377-026-02343-0}
}

@article{Rota2024EntangledQD,
  author        = {Rota, Michele B. and Krieger, Tobias M. and Buchinger, Quirin and others},
  title         = {A source of entangled photons based on a cavity-enhanced and strain-tuned {GaAs} quantum dot},
  journal       = {eLight},
  volume        = {4},
  pages         = {13},
  year          = {2024},
  doi           = {10.1186/s43593-024-00072-8}
}

@article{Chen2021PhotonicsPerspective,
  author        = {Chen, Zhigang and Segev, Mordechai},
  title         = {Highlighting photonics: looking into the next decade},
  journal       = {eLight},
  volume        = {1},
  pages         = {2},
  year          = {2021},
  doi           = {10.1186/s43593-021-00002-y}
}

@article{Li2023IntegratedMicrolaser,
  author        = {Li, Xueshi and Liu, Shunfa and Wei, Yuming and others},
  title         = {Bright semiconductor single-photon sources pumped by heterogeneously integrated micropillar lasers with electrical injections},
  journal       = {Light Sci. Appl.},
  volume        = {12},
  pages         = {65},
  year          = {2023},
  doi           = {10.1038/s41377-023-01110-9}
}

@article{Yacob2014Ripening,
author = {Yacob, M. and Reithmaier, J. P. and Benyoucef, M.},
title = {Low-density {InP}-based quantum dots emitting around the 1.5~$\mu$m telecom wavelength range},
journal = {Applied Physics Letters},
volume = {104},
number = {2},
pages = {022113},
year = {2014},
doi = {10.1063/1.4861940}
}

@article{Kors2018Ripening,
author = {Kors, Andrei and Reithmaier, Johann Peter and Benyoucef, Mohamed},
title = {Telecom wavelength single quantum dots with very small excitonic fine-structure splitting},
journal = {Applied Physics Letters},
volume = {112},
number = {17},
pages = {172102},
year = {2018},
doi = {10.1063/1.5023184}
}

\clearpage

\section{Data availability}
The data supporting the findings of this study are available from the corresponding authors upon reasonable request.

\section{Acknowledgements}
We thank Pengji Li for assistance with the measurements. This work was supported by the German Federal Ministry of Education and Research (BMBF) through the project SemIQON (13N16291) and by the European Research Council through the ERC Consolidator Grant MiNet (grant agreement No. 101043851). Funding from the Deutsche Forschungsgemeinschaft (DFG, German Research Foundation) is gratefully acknowledged under Germany's Excellence Strategy--EXC-2123--390837967, project DI 2013/6-1, and grants INST 187/704-1 FUGG, INST 187/739-1 FUGG and INST 187/740-1 FUGG. We thank the Laboratory for Nano and Quantum Engineering (LNQE) for providing access to its epitaxy and cleanroom infrastructure. M.E. gratefully acknowledges DFG support through grant INST 188/452-1 FUGG. Y.Z. acknowledges support from the China Scholarship Council (CSC; No. 201908370225).

\section{Author contributions}
F.D. conceived the project. F.D. and M.Z. supervised the research and provided resources. Y.Z. developed the CPSA growth strategy, performed and analyzed the MBE experiments, and established the surface-effective exchange dose calibration. Y.Z., Z.J., T.F. and X.C. performed and analyzed the optical measurements; Z.J. focused on the temperature-dependent PL and autocorrelation measurements. M.E. performed the HAADF-STEM and STEM-EELS measurements and carried out the corresponding analyses, including GPA. A.S. performed the ARXPS measurements; Y.Z. and A.S. analyzed and interpreted the ARXPS data. A.Ch. and C.M. fabricated the truncated-cone devices. A.Ck. performed the effective-mass calculations. D.A. contributed to literature review, discussion of the RHEED observations and the surface-effective exchange dose, and refinement of the cooling-path interpretation. E.P.R. contributed to growth-process interpretation and MBE supervision. Y.Z. wrote the original manuscript draft. M.Z. contributed to manuscript development and scientific positioning. M.Z., D.A., E.P.R., F.D. and the other co-authors reviewed and edited the manuscript. All authors discussed the results and approved the final manuscript.

\section{Competing interests}
The authors declare no competing interests.

\end{document}